\documentclass{aa}
\usepackage{graphicx}

\begin{document}

\sloppypar

%
   \title{First simultanous 
X-ray and optical observations of rapid variability of supercritical accretor SS433}

   \author{M. Revnivtsev \inst{1,2}, R. Burenin \inst{2}, 
   S. Fabrika \inst{3}, K. Postnov\inst{4}, \\  I.Bikmaev \inst{5}, 
   M. Pavlinsky \inst{2}, R.Sunyaev \inst{1,2}, I. Khamitov \inst{6}, Z.Aslan \inst{6}}

   \offprints{mikej@mpa-garching.mpg.de}

   \institute{
              Max-Planck-Institute f\"ur Astrophysik,
              Karl-Schwarzschild-Str. 1, D-85740 Garching bei M\"unchen,
              Germany,
	  \and   
              Space Research Institute, Russian Academy of Sciences,
              Profsoyuznaya 84/32, 117810 Moscow, Russia
         \and   
              Special Astrophysical Observatory, Nizhnij Arkhyz, 
Karachaevo-Cherkesiya, 369167,  Russia
         \and   
              Sternberg Astronomical Institute, 119992, Moscow, Russia
         \and   
              Kazan State University, Kremlevskaya
            str.18, 420008, Kazan, Russia
	\and 
	       TUBITAK National Observatory, Turkey
            }
  \date{}

        \authorrunning{Revnivtsev et al.}
       \titlerunning{Simultaneous X-ray and optical variability of SS433}
 
   \abstract{We present results of first simultaneous optical and X-ray 
observations of peculiar binary system SS433. 
For the first time, chaotic variability of SS433
in the optical spectral band ($R$ band) on time scales as small as tens of 
seconds was detected.
We find that the X-ray
flux of SS433 is delayed with respect to the optical emission by 
approximately 80 sec. Such a delay can be interpreted as the 
travel time of mass accretion rate perturbations  
from the jet base to the observed X-ray emitting region. 
In this model, the length of the supercritical accretion disk funnel
in SS433 is $\sim 10^{12}$ cm.
   }

   \maketitle

%

\section{Introduction}

SS433 is the unique X-ray binary, demonstrating precessing jets and
supercritical accretion disk around a black hole (Margon, 1984; see 
Fabrika, 2004 for a recent review). The total luminosity
of the system is estimated to be at a level of $10^{40}$ erg/s,
with most of this luminosity coming in UV and optical bands (Cherepashchuk
et al., 1982; Dolan et al., 1997). 
According to the current model of this binary, 
the dense outflowing wind and the geometrically thick 
accretion disk totally cover from us the direct 
X-ray emission from the 
innermost hot regions of the accretion disk. The early observations 
of SS433 by EXOSAT and {\it ASCA} satellites led to the conclusion that 
standard X-ray emission in SS433 originates in optically thin plasma
outflowing in a mildly relativistic, $v=0.26c$, jets (Watson et al. 1986, 
Kotani et al. 1996). 
A detailed analysis of high-resolution {\it Chandra}
observations of SS433 suggests that the X-ray jet consists of well collimated
freely expanding cooling plasma with temperature varying from 20 to 0.5 keV.
The size of the vizible X-ray jet is estimated to be $\sim 10^{10}-10^{11}$
cm (Marshall et al. 2002). This is the lower limit on the size of X-ray
emitting jet 
if heating effects are not important (Brinkmann and Kawai 2000).
The observed X-ray luminosity ($\approx 10^{36}$ erg/s) is only a tiny part 
of the total jet power. For example,
the kinetic power of the jet is estimated to be 1000 times larger
than the observed X-ray luminosity. The contribution of the observed 
X-ray luminosity
to the total energetics of the system is even smaller -- not more than 
$10^{-4}$.

The long-term behavior of SS433 has been well studied at different
wavelengths. The source demonstrates different types of long-term 
periodicities: precessional (162.3 days), orbital (13.082 days), and
nutational 
(6.28 days) ones. SS433 exhibits noticeable 
variability on shorter times as well,
however at time scales smaller than hours this has been poorly studied. 
For example, erratic variability with an amplitude of 5-10\,\% on a
time-scale of a few minutes was found 
in the optical band by Goranskii et al. (1987) and Zwitter et al. (1991).
Recently, Kotani et al. (2002) detected X-ray variability of SS433
(RXTE, PCA) on a time scale of $\sim 50$~s, when the source was 
in its active state.

The accretion disk can modulate energy release and luminosity in the
broad range of time scales (\cite{lyubarsky97}). This mechanism can nicely
explain the observed flicker-noise spectra found in
some X-ray binaries, e.g.
Cyg X-1 (\cite{chur01}). In the case of SS433 
the X-ray emission of the accretion disk
is not directly visible, so variability in the optical and UV 
may be a better indicator of the temporal properties of the 
accretion disk. 


The UV and optical radiation of SS433 is well approximated by a single
reddened black body source (Cherepashchuk et al., 1982; Dolan et al., 1997;
Gies et al., 2002) with a temperature $T_e \sim 5 \cdot 10^4$~K
($A_V \approx 8$) and a size $r_{ph} \sim (1-2) \cdot 10^{12}$~cm. 
The bulk of the optical and UV emission most likely escapes
from the hot  funnel in the photosphere close to
the jet bases (Dolan et al. 1997; Fabrika 2004). 
Should X-ray and optical variabilities be 
due to one physical reason (e.g., the accretion rate modulation),  
simultaneous observations of SS433 in the optical band 
(which reflect variability of the energy release in the 
supercritical accretion disk) and in X-rays
(that come from the footpoint of the jet) will be invaluable 
to test the origin of jets in SS433 and their close surroundings.  


In this paper we
present the analysis of simultaneous optical and X-ray observations of
SS433 in March 2004. 

\section{Observations and data analysis}

\subsection{RTT150}
Optical observations of SS433 were performed with 1.5-m Russian-Turkish
Telescope (RTT150) at T\"{U}BITAK National Observatory (TUG), Bakyrly
mountain, 2547\,m, Turkey, $2^\mathrm{h}01^\mathrm{m}20^\mathrm{s}$~E,
$36^\circ49'30''$~N. The object was observed as a target of opportunity, in
order to support INTEGRAL and RXTE observations in optical.

Observations were performed on March 25 (0:33--3:08\,UT) and March 27
(0:18--3:20\,UT) under clear sky and poor seeing ($\approx2''$) conditions.
We used low readout noise back-illuminated 2$\times$2\,K Andor Technologies
DW436 CCD, mounted in F/7.7 Cassegrain focus of the telescope.  In our
observations we used R filter because the Galactic absorption in R is
significantly lower than in blue bands and SS433 is quite bright in R
(R$\approx 12$).

To improve the temporal resolution of our measurements we used only a small
CCD subframe which contains small part of the field near the object. Also we
used 4$\times$4 binning mode and high readout speed mode of the CCD. Time
resolution $\approx3.4$\,s with 1\,s exposure was obtained. The data
reduction was done with IRAF (Image Reduction and Analysis
Facility)\footnote{http://tucana.tuc.noao.edu/} and using our own software.
Bias and dark counts were subtracted and flat field correction applied for
all the images. Apart of SS433 our subframe contains few bright stars which
were used as reference for differential photometry. 
All stars were centered only once, using one
reference image. Centers of stars in other images were calculated from their
shifts, therefore centering errors (if any) are exactly the same in object
and reference stars.

\subsection{RXTE}

X-ray observations of SS433  were performed by the RXTE observatory 
(\cite{rxte}) on March 25 and 27 in respond to the TOO request.
The total exposure time of two observations amounts to about 4 ks.
In order to study the variability of the source we used primarily
the Proportional Counter Array (PCA) data, which have large effective area.
Data reduction of the RXTE/PCA data was done using standard tasks of 
the LHEASOFT/FTOOLS 5.2 package. In order to improve the statistics
of the lightcurve, we used data from all operational detectors but 
PCU0, which lacks propan veto layer. As the source is relatively weak, 
for the background modelling we accepted CMl7\_240 model.

\section{Results}

During our observations, SS433 was observed in  
the orbital elongation (March 25) and
the superiour conjunction of the accretion disk (March 27), while 
the precessional phase was close to the most open disk. 
Both binary and precessional phases are most
favourable for studying the base of the blue jet. Radio monitoring of 
SS433 indicated (Trushkin 2004) that 
in the beginning of March, 2004, the source entered into 
its active state with several strong flares on March 10-16, 2004. 

\begin{figure*}
\hbox{
\includegraphics[width=\columnwidth,bb=33 176 570 715,clip]{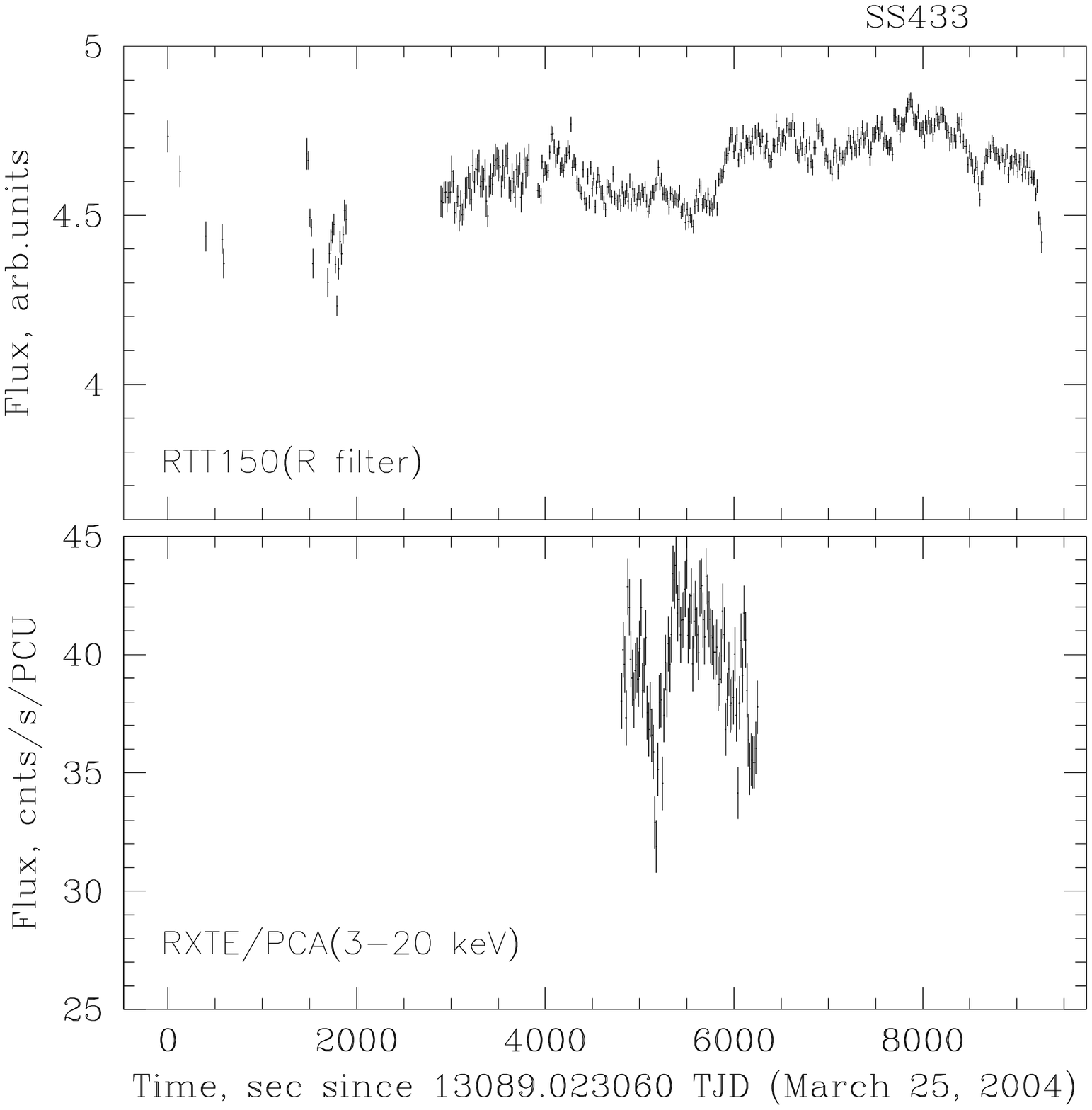}
\includegraphics[width=\columnwidth,bb=33 176 570 715,clip]{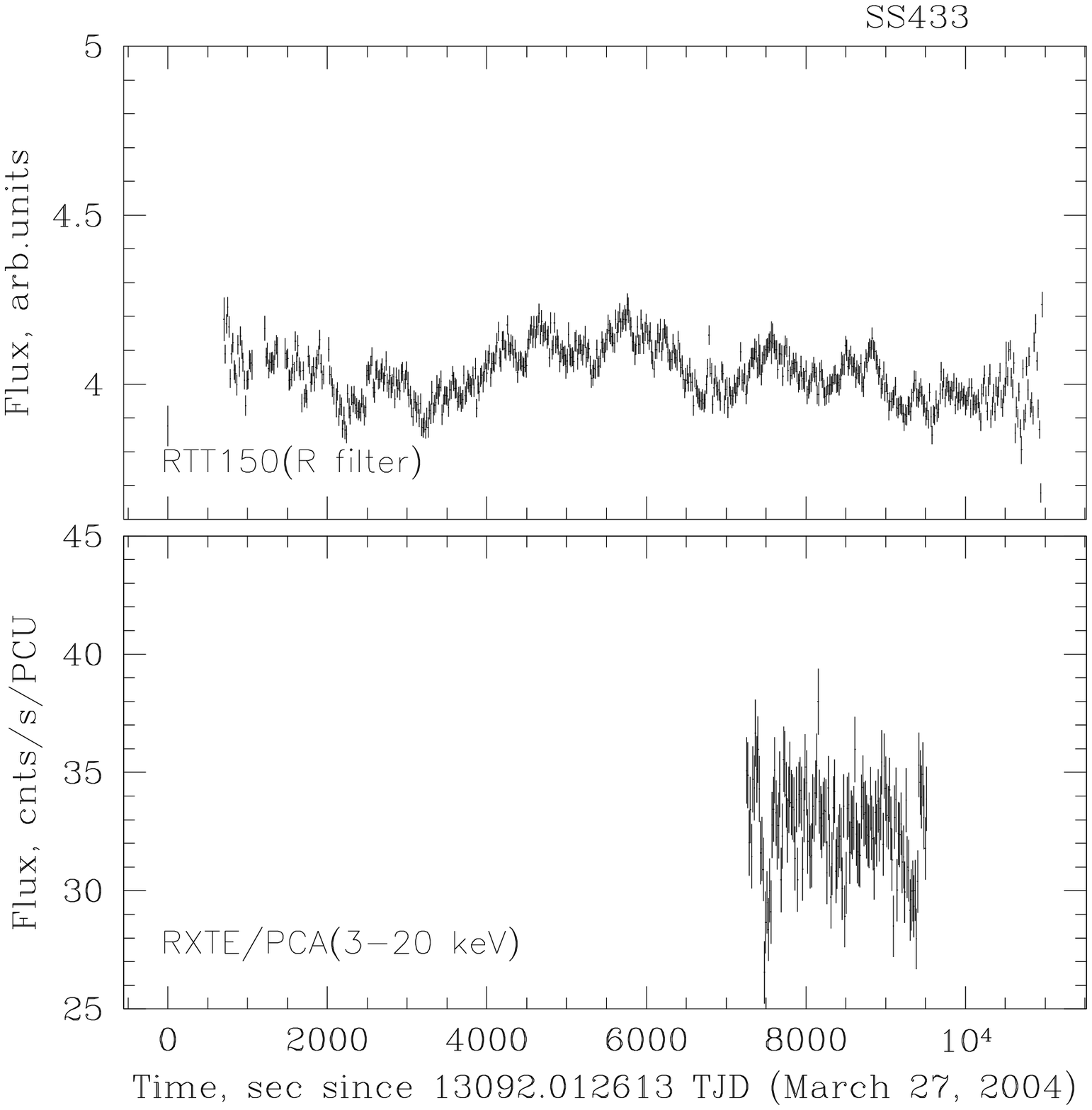}
}
\caption{Lightcurves of SS433 in the optical (RTT150, $R$-band) and X-ray 
(RXTE/PCA, 3-20 keV) energy bands.\label{lcurves}}
\end{figure*}

\begin{figure}[htb]
\includegraphics[width=\columnwidth,bb=33 186 570 715,clip]{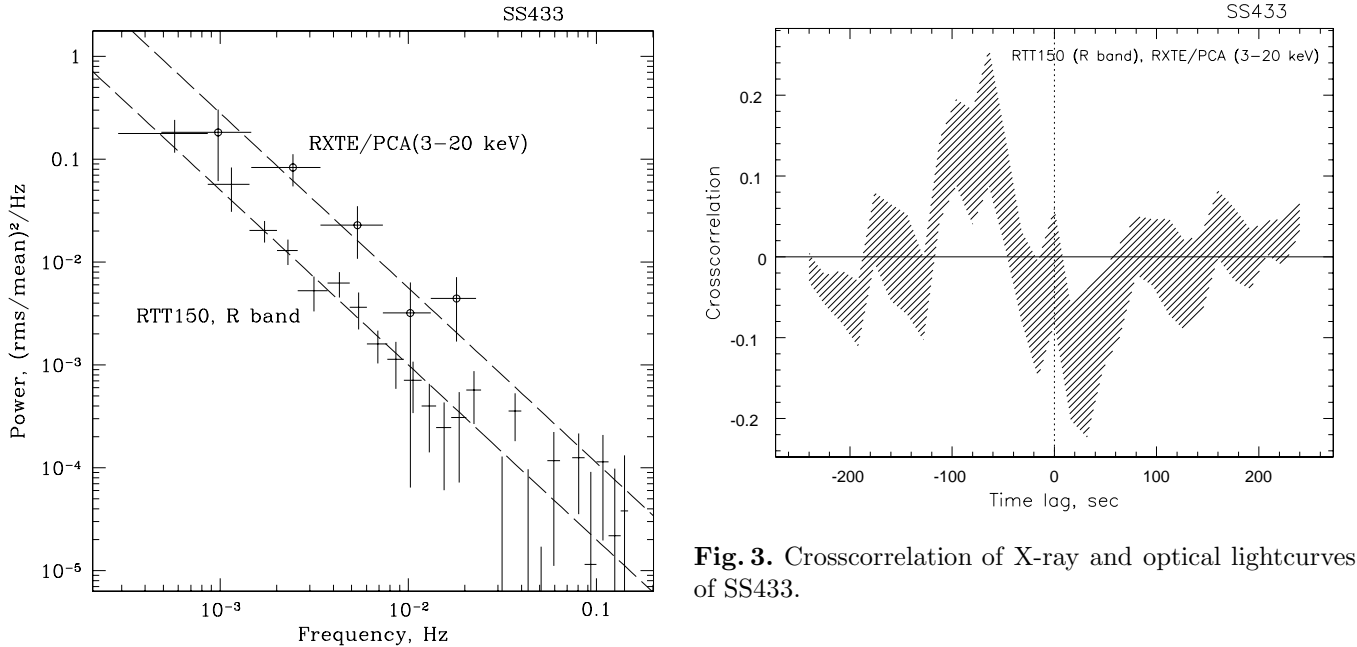}
\caption{Power spectra of the optical and X-ray  
lightcurves of SS433. The dashed line is the best fit 
model to the optical power spectrum. For X-ray points this power law was 
rescaled.\label{power}}
\end{figure}

Two sets of simultanous optical and X-ray measurements of the flux of SS433
by RTT150 and RXTE were obtained.
The observed lightcurves are presented in Fig.\ref{lcurves}. During both 
observations chaotic variability is clearly visible in both optical 
and X-ray lightcurves. The power spectrum of this variability can be
described by a simple power law from frequencies where the statistical
noise dominates ($\sim$0.05--0.1 Hz) down to $10^{-4}$ Hz. The slope
of the power law in the optical band ($R$ filter) is $\alpha=-1.7\pm 0.1$ (see 
Fig.\ref{power}). The same power law can also describe the power spectrum 
obtained from X-ray data.
In the frequency band $5\times10^{-4} - 0.01$ Hz,  
the amplitude of variability
is approximatety $1.5\pm0.2$\% and $6.5\pm1.4$\% 
for optical and X-ray lightcurves, respectively. 

\begin{figure}[htb]
\includegraphics[width=\columnwidth,bb=40 180 580 608, clip]{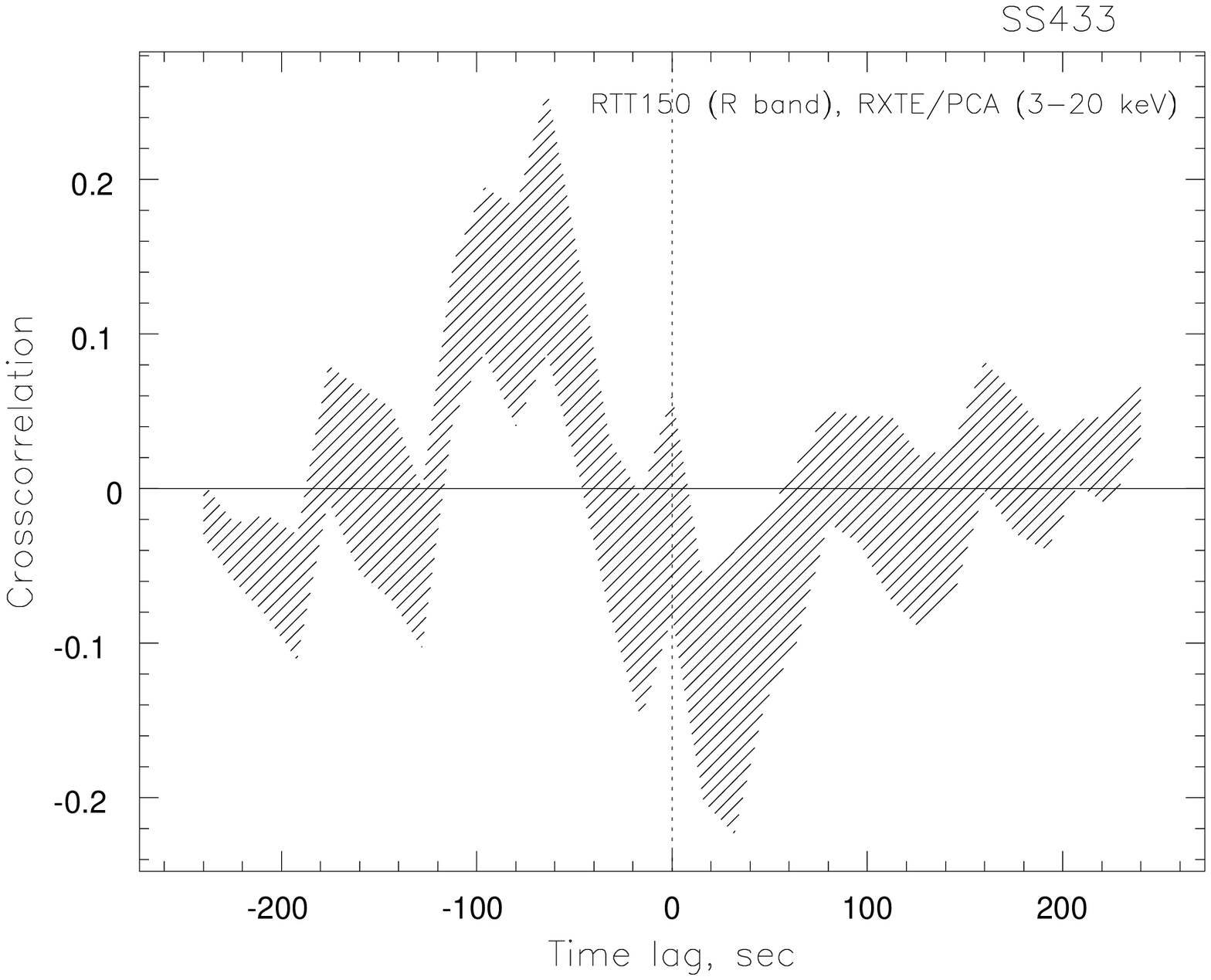}
\caption{Crosscorrelation of X-ray and optical lightcurves of SS433.\label{crosscor}}
\end{figure}

The statistically significant chaotic source variability in both optical and 
X-ray band allow us to study their crosscorrelation. It is a big advantage 
for us that the lightcurves of SS433 do not demonstrate detectable 
quasiperiodical  variability, because in this case interpretation of the 
crosscorrelation picture would be ambiguous.

The relatively high 
fractional stochastic variability of the optical and X-ray flux 
enabled us to carefully check 
the stability of the crosscorrelation pattern. To this end, 
we split all overlaping lightcurves into 256-sec intervals and inside each
interval calculated X-ray-to-optical crosscorrelation. 
Then we averaged the obtained crosscorrelations and calculated the root 
mean square deviations of crosscorrelations in individual intervals 
from the average one.
The resulted average crosscorrelation is presented in Fig.\ref{crosscor}. 
The dashed region shows the rms deviations of
individual  
crosscorrelations from the average one. One can clearly see from the
plot that the X-ray lightcurve lags the optical one (negative time lag 
means that the optical variability preceeds the X-ray one). 

\section{Discussion}

Simultaneous observations
of SS433 in optical and X-ray energy bands demonstrate clear
correlation of measured fluxes. The variable part of the 
X-ray lightcurve is delayed with 
respect to the optical one by about 80 sec. 

The lag of X-rays with respect to the optical variability can be
anticipated in the framework of the cooling jet model that appears
as a result of acceleration of matter in the base of the funnel 
in the center of the supercritically accrettion disk (Calvani, Nobili 1981;
Bodo et al. 1985; Eggum et al. 1988; Okuda 2002). 

The variability of SS433 on longer time scales with lagged correlation   
has been known 
in different parts of the emitting regions of this system
(Cherepaschuk 2002, Fabrika 2004).
For example, the optical nutation variability has been observed to 
preceed the nutation motion of optical jets by 
0.6 days. This is exactly the time it takes for the 
jet gas to travel from the compact 
object to the $H\alpha$ emitting region downstream 
the jets. A maximum of $H\alpha$ emission in the optical jets 
forms at a distance of $4 \cdot 10^{14}$cm from the source (Borisov and Fabrika 1987). 
The amplitude of the nutation variability (Cherepaschuk 2002) is 
$\Delta V \approx 0.17$ and increases in shorter $B$ band. This
nutational lag proves that the optical and UV radiation
come from the inner wind region, i.e. from the jet base, but not from
the jet itself. The jets' energy budget ($L_{kin}\sim 10^{39}$~erg/s,
$L_{x}\sim 10^{36}$~erg/s) is insufficient to provide the optical nutational
modulation, which is 20~\% of $L_{opt}\sim 10^{38}$~erg/s, 
$L_{UV}\sim 10^{40}$~erg/s.

In our case we are dealing with two nearby emitting regions - the innermost
region of the supercritical accretion disk, which emits most of the energy
in the optical and UV band, and the X-ray emitting jet, which is supposed to
be launched near the compact object and have a bulk velocity of $0.26c$. The
most straightforward and simplest interpretation of the observed time delay
between the optical and X-ray fluxes, which does not contradict the common
knowledge about SS433, can be as follows.

Mass accretion fluctuations appear in the bottom of the supercritical
disk funnel where the jet is being formed  
and results in fluctiating X-ray flux
inside the funnel. According to the supercritical disk simulations (Eggum et
al. 1988; Okuda 2002), the place of the jet formation (10-100~$r_g$) is quite 
close to the black hole. The X-ray radiation coming upwards the funnel may be
absorbed by the outer funnel walls (which are 
directly observed or lie immediately close to the directly
observed parts of the funnel) 
and produce fluctuating UV-optical radiation via photoabsorption.
Similar processes are well-known to operate in low mass X-ray binaries
(see e.g. \cite{lawrence83}).

The same fluctuations reach the observed part of the X-ray jet after
the jet propagation time inside the funnel and may appear as 
variable fraction of the visible X-ray emission. 
This is the basic
mechanism we propose to explain the correlated X-ray - optical variability.
It is important that this correlated variability between
optical and X-ray variations has to be related to all time-scales generated in
the accretion disk, because we do not see a QPO-like variability.

The delayed X-ray variability allows us to estimate the funnel length 
$r_f$ in SS433. The time it takes for the X-ray jet to appear above the
funnel edge is $r_f/V_j$, the time for the light to reproduce optical
emission via photoabsorption of X-rays from the funnel bottom
is $\eta r_f/c$, where factor 
$\eta \ga 1$ is responsible for possible delays (geometry, scattering, etc.). 
The difference between these
two times yields the time delay between the optical and X-ray variabilities
found by us, $\Delta t \sim 80$~sec. So for the funnel length we have 
the estimate 
$r_f \sim 2 \cdot 10^{12}/(c/V_j-\eta)$~cm.  
A lower limit on the funnel length is for $\eta=1$, i.e. $r_f \ga 10^{12}$~cm. 
So the bases of the X-ray jets, 
whose length $\ga 10^{10}-10^{11}$~cm were derived from 
the {\it Chandra}  spectra 
(Marshall et al. 2002), have to be placed at distance $r_f\sim 10^{12}$ cm 
from the black hole.

\begin{acknowledgements}
The authors thank S.F.Trushkin for drawing their attention to the activity
of SS433 in March 2004, to A.I.Zakharov for helpful discussions.
The authors are grateful to the 
RXTE planning team for rapid responce to the TOO request.
This work was partially supported by grants of
Minpromnauka NSH-2083.2003.2, NSH-1789.2003.02 and program of Russian
Academy of Sciences ``Non-stationary phenomena in astronomy''.
Partial support through RFBR grants 03-02-16110, 04-02-16349 and 02-02-17174
is acknowledged. MR, RB, IB and MP thank International Space Science 
Institute (ISSI, Bern) for partial support.
Research has made use of data obtained from 
High Energy Astrophysics Science Archive Research Center 
Online Service, provided by the NASA/Goddard Space Flight Center.

\end{acknowledgements}

\end{document}